\documentclass[conference]{IEEEtran}
\IEEEoverridecommandlockouts
\usepackage{cite}
\usepackage{amsmath,amssymb,amsfonts}
\usepackage{algorithmic}
\usepackage{graphicx}
\usepackage{textcomp}
\usepackage{xcolor}
\usepackage{url}
\usepackage{array}
\usepackage{siunitx}
\usepackage{comment}
\usepackage{hyperref}
\usepackage[capitalise]{cleveref}
\usepackage{etoolbox}
\usepackage{booktabs}
\usepackage{subfig}
\usepackage{balance}

\usepackage{fancyhdr}

\newcommand{\ubold}{\fontseries{b}\selectfont}
\robustify\ubold

\def\BibTeX{{\rm B\kern-.05em{\sc i\kern-.025em b}\kern-.08em
    T\kern-.1667em\lower.7ex\hbox{E}\kern-.125emX}}

\makeatletter
\newcommand{\linebreakand}{%
  \end{@IEEEauthorhalign}
  \hfill\mbox{}\par
  \mbox{}\hfill\begin{@IEEEauthorhalign}
}
\makeatother

\usepackage{eso-pic}
\usepackage{url}
\AddToShipoutPictureBG*{
  \AtPageUpperLeft{%
    \put(0,-40){\raisebox{15pt}{\makebox[\paperwidth]{\begin{minipage}{21cm}\centering
      \textcolor{gray}{This article has been accepted for publication in the proceedings of the \\2024 IEEE International Conference on Omni-layer Intelligent Systems (COINS). DOI: 10.1109/COINS61597.2024.10622644} 
    \end{minipage}}}}%
  }
  \AtPageLowerLeft{%
    \raisebox{25pt}{\makebox[\paperwidth]{\begin{minipage}{21cm}\centering
      \textcolor{gray}{ \copyright 2024 IEEE.  Personal use of this material is permitted.  Permission from IEEE must be obtained for all other uses, in any current or future media, including reprinting/republishing this material for advertising or promotional purposes, creating new collective works, for resale or redistribution to servers or lists, or reuse of any copyrighted component of this work in other works.
      }
    \end{minipage}}}%
  }
}

\begin{document}

\title{Noise Analysis and Modeling of the PMD Flexx2 Depth Camera for Robotic Applications}

\author{\IEEEauthorblockN{
      Yuke~Cai\IEEEauthorrefmark{1},
      Davide Plozza\IEEEauthorrefmark{1},
      Steven~Marty\IEEEauthorrefmark{1},
      Paul~Joseph\IEEEauthorrefmark{1},
      Michele~Magno\IEEEauthorrefmark{1}}

      \IEEEauthorblockA{\IEEEauthorrefmark{1}ETH Z\"{u}rich, Switzerland}
      \IEEEauthorblockA{\{yukcai,dplozza,martyste,josephp,magnom\}@ethz.ch}
      }

\maketitle

\begin{abstract}
Time of Flight (ToF) cameras, renowned for their ability to capture real-time 3D information, have become indispensable for agile mobile robotics. These cameras utilize light signals to accurately measure distances, enabling robots to navigate complex environments with precision. Innovative depth cameras, characterized by their compact size and lightweight design, such as the recently released PMD Flexx2, are particularly suited for mobile robots. Capable of achieving high frame rates while capturing depth information, this innovative sensor is suitable for tasks such as robot navigation and terrain mapping. Operating on the ToF measurement principle, the sensor offers multiple benefits over classic stereo-based depth cameras.
However, the depth images produced by the camera are subject to noise from multiple sources, complicating their simulation.
This paper proposes an accurate quantification and modeling of the non-systematic noise of the PMD Flexx2. 
We propose models for both axial and lateral noise across various camera modes, assuming Gaussian distributions. Axial noise, modeled as a function of distance and incidence angle, demonstrated a low average Kullback–Leibler (KL) divergence of 0.015 nats, reflecting precise noise characterization. Lateral noise, deviating from a Gaussian distribution, was modeled conservatively, yielding a satisfactory KL divergence of 0.868 nats. These results validate our noise models, crucial for accurately simulating sensor behavior in virtual environments and reducing the sim-to-real gap in learning-based control approaches.

\end{abstract}

\begin{IEEEkeywords}
Depth sensor, ToF, PMD Flexx2, noise model, noise characterization, robot perception.
\end{IEEEkeywords}

\section{Introduction}\label{sec:intro}

In recent years, depth cameras have been increasingly applied across various fields\cite{mejia2019kinect, el2012study, hutter2012starleth, pinto2015evaluation, zennaro2014evaluation, miki2022learning, galna2014accuracy}. In robotics, these cameras serve as essential components of the perception systems. They are particularly important in legged robots, where they provide depth information for locomotion control and obstacle detection\cite{miki2022learning}. Furthermore, the integration of depth cameras enables the robots to leverage exteroceptive information, like environmental height maps, for enhanced navigation and control \cite{el2012study, hutter2012starleth, pinto2015evaluation, zennaro2014evaluation, miki2022learning}. 

Depth cameras fall into three categories such as stereo imaging, structured light or Time of flight (ToF)~\cite{funek2019evaluation}. 
Stereo cameras such as the Realsense D435 are widely used~\cite{miki2022learning} and their noise models have been well characterized~\cite{ahn2019analysis}.
Existing ToF such as the Microsoft Kinect~\cite{fankhauser2015kinect,khoshelham2012accuracy,wasenmuller2017comparison} or PMD Camboard
Pico Flexx~\cite{pasinetti2019performance} have similarly undergone thorough evaluation and comparison.
The novel PMD Flexx2 sensor \cite{pmdflexx2}, an improved version of the PDM Camboard Pico Flexx, offers multiple benefits to its predecessor in terms of measurement range and achievable frame rate.

Despite their utility in the robotics field, these devices exhibit inherent imperfections and are prone to noise, which can be influenced by a variety of factors including operating principles, measurement distance, object material properties, and external lighting conditions~\cite{funek2019evaluation,haider2022cameranoise}. These limitations not only compromise the accuracy and reliability of the data obtained but also hinder the broader adoption and optimal utilization of depth cameras in robotics.

Denoising algorithms are proposed in recent literature proposing interesting solutions to mitigate such issues, enhancing the robustness of algorithms reliant on depth images ~\cite{yan2020depth}.
Alternatively, a more comprehensive and innovative approach involves understanding and modeling the noise characteristics intrinsic to these cameras\cite{mallick2014characterizations,ahn2019analysis}.
This understanding is beneficial for improving perception systems, such as 3D reconstruction~\cite{nguyen2012modeling} or mapping algorithms~\cite{ahn2019analysis}.
Furthermore, noise models are crucial for accurate simulation models.
Such models are also vital to the successful deployment of Reinforcement Learning (RL) algorithms that make use of those sensors, which are often initially trained and tested in simulated environments  \cite{miki2022learning, margolis2023walk,  lee2020learning}.
In these cases accurately replicating real-world sensor noise in simulations, particularly from depth cameras, is crucial for seamless sim-to-real transfer \cite{gervet2023navigating}. 

\begin{figure}[t]
    \centering
    \includegraphics[width=0.85\linewidth]{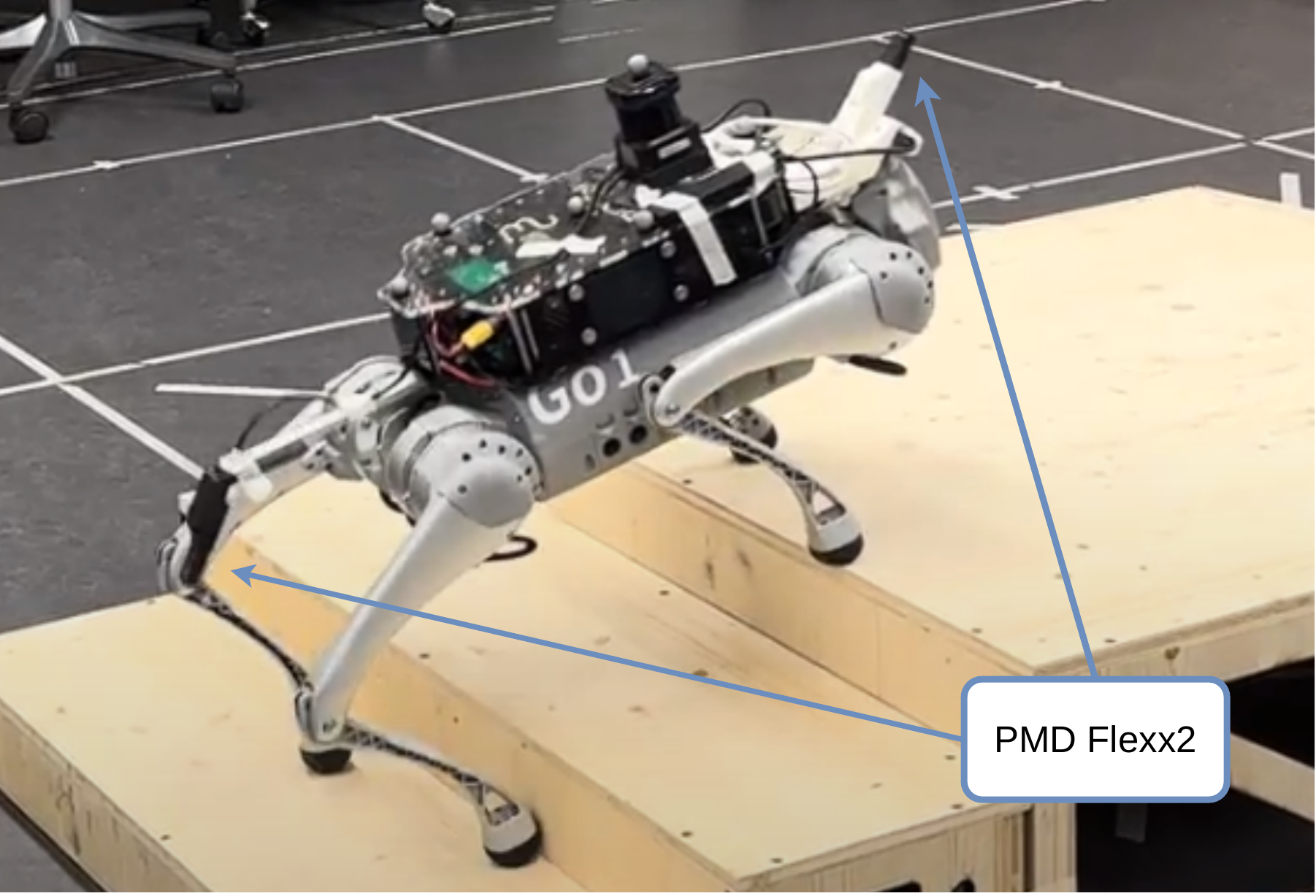}
    \caption{Quadrupedal robot equipped with two PMD Flexx2 cameras.}
    \label{fig:robodog}
\end{figure}

This paper focuses on the characterization and modeling of the non-systematic noise in the PMD Flexx2 depth camera.
Through this endeavor, we aim to provide a foundational noise model that improves the accuracy of the sensor simulation in virtual environments.

\begin{figure}[t]
    \centering
    \includegraphics[width=0.65\linewidth]{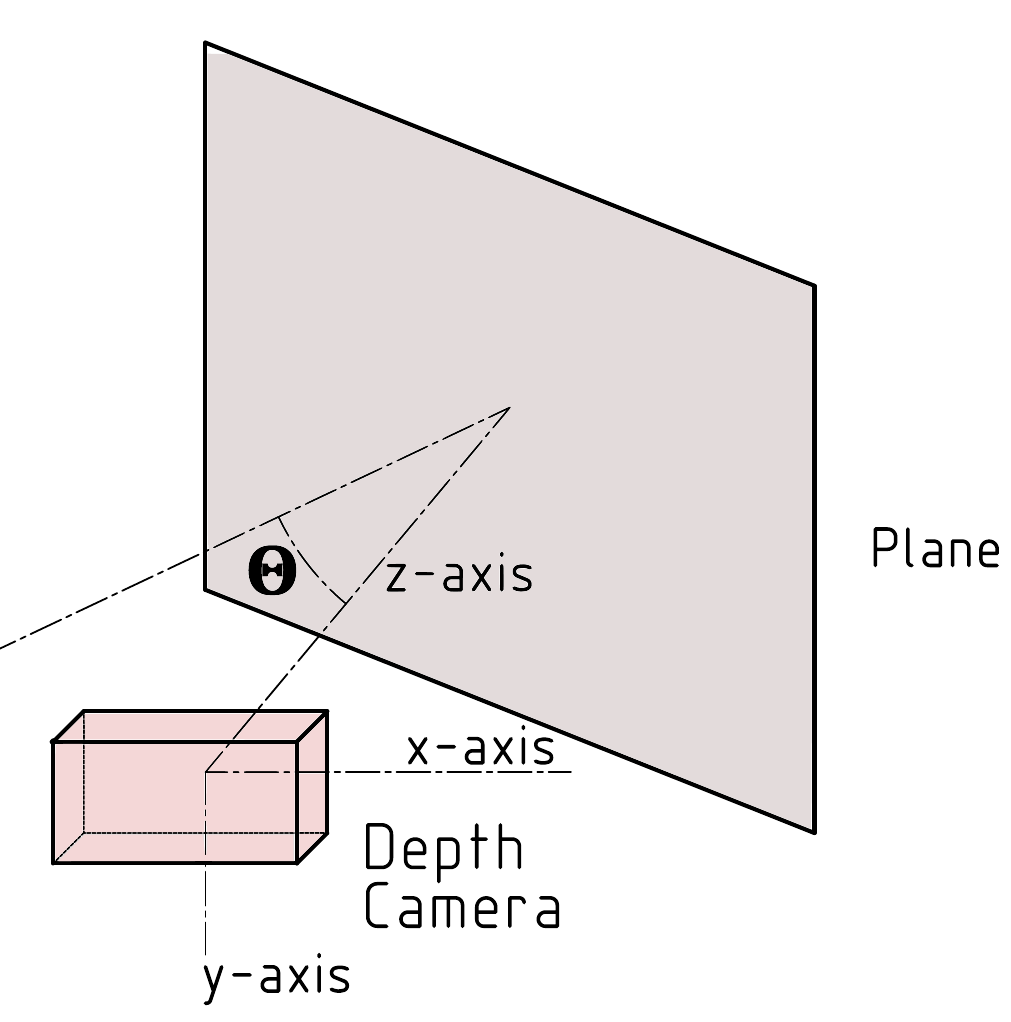}
    \caption{Testing environment setup, where $\Theta$ is the incident angle.}
    \label{fig:env_setup}
\end{figure}

\section{Related Works}\label{sec:related}

In the evolving landscape of mobile robotics, a diverse array of depth cameras have been deployed to capture depth images. The quest for high-quality depth images necessitates ongoing research into the characterization and modeling of the inherent noise in these cameras' measurements. A previous study on the Kinect camera~\cite{khoshelham2012accuracy} has underscored the correlation between the random noise in-depth measurements and the increasing distance from the object, though this study omitted considerations of the incidence angle's impact. To measure and investigate the noise, they measured a planar surface of a door at various distances and analyzed the acquired depth images.

\begin{table}[b]
    \caption{Comparison between different Depth Cameras and the Noise Evaluations that have been done}
    \begin{center}
    \begin{tabular}{lcccc}
    \toprule
    Model              & \ubold Kinect V2     & \ubold D435      &  \ubold Pico Flexx & \ubold Flexx2         \\ \midrule
    Max. Res. & 512 x 424     & 1280 x 720  & 224 x 171   & 224 x 172          \\ 
    Angle (°)      & 70.6 x 60     & 87 x 58  &  62 x 45  & 56 x 44            \\ 
    Max. FPS & 30            & 90      &   45   & 60                 \\ 
    Weight (g)          & 970           & 75     &  8     & 13                 \\ 
    Dim. (mm)     &249×66×67&90x25x25&68x17x7 &72x19x11\\ 
    Range (m)  & 0.5–4.5           & 0.3-3     &  0.1-4  & 0.1-7    \\ 
    Principle  & ToF & Stereo & ToF & ToF\\
    Evals. & \cite{fankhauser2015kinect,zennaro2014evaluation,pinto2015evaluation} & \cite{ahn2019analysis} & \cite{pasinetti2019performance}& -\\
    \bottomrule
    \end{tabular}
    \end{center}
    \label{tab:camera_compare}
\end{table}

Further investigations on the Kinect camera have led to the development of methodologies aimed at modeling the noise specific to depth cameras, offering insights that are potentially applicable across various models\cite{nguyen2012modeling, mallick2014characterizations}.
Notably, Nguyen et al.\cite{nguyen2012modeling} constructed an intricate noise model for the Kinect sensor by modeling the noise separately as axial and lateral distribution models, each as a function of the distance and angle of incidence. This approach inspired our method, where we similarly separate and model the noise components.

With the advent of Kinect V2, detailed in \autoref{tab:camera_compare}, some works have investigated the error and noise of this sensor as well as factors which can influence the measurement\cite{breuer2014low,lachat2015first, pinto2015evaluation, wasenmuller2017comparison}.
Fankhauser et al.\cite{fankhauser2015kinect} have contributed to the formulation of an empirical model that delineates noise behavior as a function of both the distance and the angle of the observed surface, under indoor and outdoor conditions.
Similarly, to obtain depth images for analysis, they placed a rotatable planar surface in front of Kinect V2 sensor and measured at different distances and various angles.

The introduction of the Intel RealSense D400 series spurred additional research on the noise modeling of these cameras\cite{giancola2018metrological, ahn2019analysis}. Ahn et al.\cite{ahn2019analysis} have successfully analyzed and modeled the noise of the Intel RealSense D435, further contributing to the field's collective knowledge base. They utilized a similar approach to \cite{fankhauser2015kinect} and \cite{nguyen2012modeling} to obtain depth images of a planar surface for the modeling of noise.

Transitioning from bulky ToF cameras to smaller, lightweight options, PMD introduced the Camboard Pico Flexx. As indicated in \autoref{tab:camera_compare}, its size falls within the range of the D435 stereo camera, with increased FPS compared to the Kinect V2.
Pasinetti et al.~\cite{pasinetti2019performance} showed how despite its smaller size, the Pico Flexx can achieve comparable results to the Kinect V2. 
Further studies on the Pico Flexx compare and evaluate the noise for applications as agricultural application~\cite{bahnsen20213d} or Sewer Inspection~\cite{condotta2020evaluation}.
These works only evaluate the camera for a specific use cases rather than providing a generalized noise modeling approach.
The newest ToF camera from PMD, namely the Flexx2 offers a bigger range and higher FPS compared to its predecessor as seen in \autoref{tab:camera_compare}.
This makes it an ideal candidate for integration into mobile robots. However, a gap in the literature is evident, as the noise characteristics of the PMD Flexx2 have yet to be extensively studied and modeled.

This work aims to bridge this gap by presenting an analysis and empirical modeling of the noise characteristics specific to the PMD Flexx2 sensor.
To the best of the authors' knowledge, this research represents the first comprehensive attempt to characterize and quantify the noise performance of the PMD Flexx2, providing crucial insights for its application in complex robotic systems.

\begin{figure}[t]
  \centering

  \subfloat[]{\includegraphics[width=0.45\linewidth]{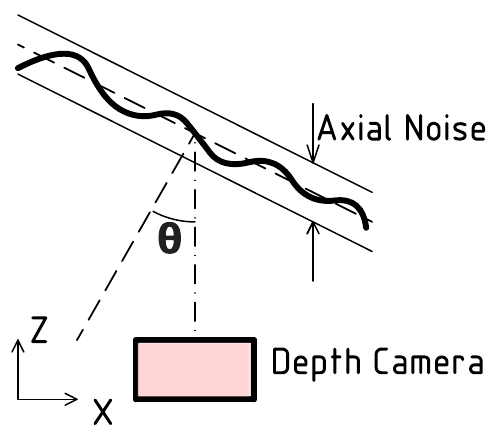}\label{fig:axial_noise}}
  \hfill
  \subfloat[]{\includegraphics[width=0.45\linewidth]{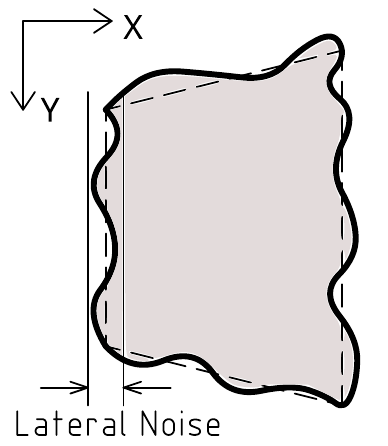}\label{fig:lateral_noise}}

  \caption{ \ref{fig:axial_noise}: A top-down cross-section of the depth map, \ref{fig:lateral_noise}: 2D projected depth map of a planar target.}
  \label{fig:noise_model}
\end{figure}

\section{System Setup}\label{sec:system}

This paper is motivated by the development of an exteroceptive RL controller for quadrupedal robotic locomotion. This controller leverages depth perception enabled by two PMD Flexx2 cameras mounted on the Unitree Robotics Go1 quadrupedal robot depicted in \autoref{fig:robodog}.
Initially trained in simulated environments, these controllers face significant challenges when transitioning to real-world deployment due to discrepancies in sensor data caused by noise. Accurate characterization and modeling of the depth sensor noise are thus crucial. We refine our simulation's fidelity through precise noise modeling of the PMD Flexx2 sensor, intending to enhance the controllers' performance and reliability in actual operating conditions, thereby reducing the sim-to-real gap \cite{gervet2023navigating}.

\subsection{PMD Flexx2 3D Camera}\label{sec:spec}

\begin{figure}[t]
  \centering
  \subfloat[]{\includegraphics[width=0.50\linewidth]{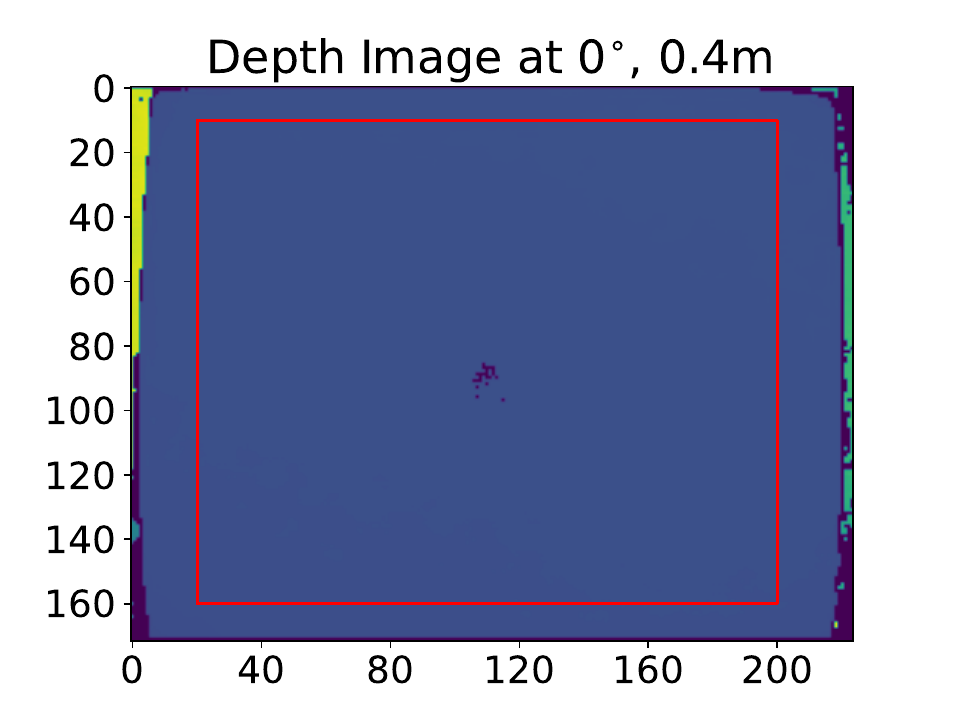}\label{fig:axial_box_0}}
  \hfill
  \subfloat[]{\includegraphics[width=0.50\linewidth]{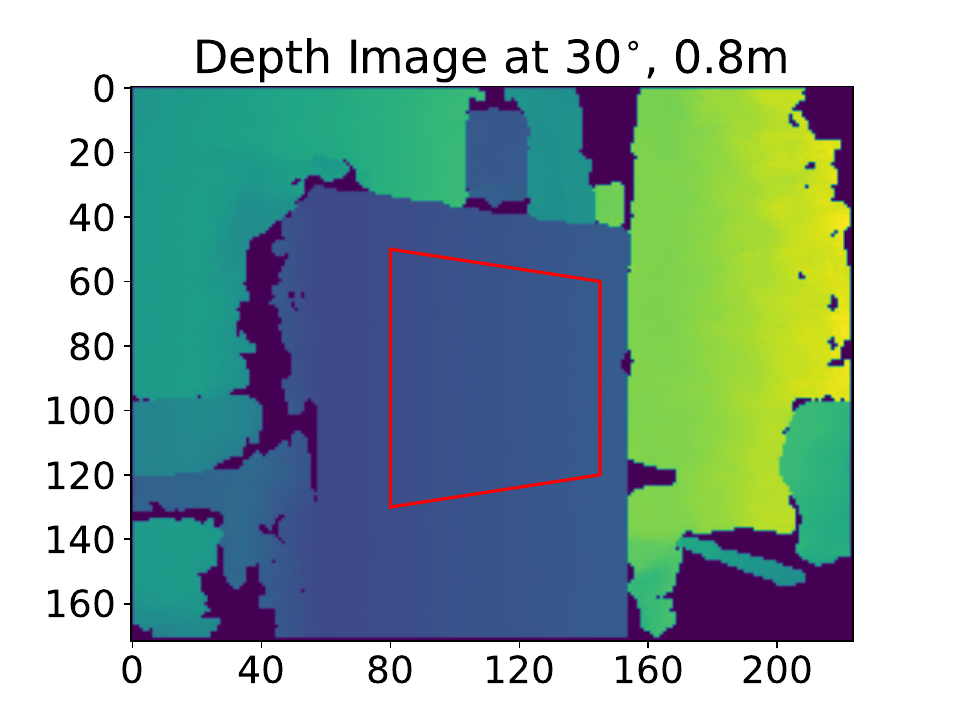}\label{fig:axial_box_1}}
  \caption{Examples of depth images with region chosen for axial noise analysis.}
  \label{fig:axial_box}
\end{figure}

\begin{table}[bp]
    \caption{Measured Operation Modes of PMD Flexx2}
    \begin{center}
    \begin{tabular}{cccc}
    \toprule
    \textbf{Frequency Pairs}& \textbf{Mode Name}& \textbf{Frame Rate}& \textbf{Range} \\
    \midrule
    \qty{60}{Mhz} & \textit{Mode\_5\_15fps} & \qty{15}{Hz}& \qty{0.1}{m}-\qty{2.4}{m} \\
    \qty{60}{Mhz} & \textit{Mode\_5\_30fps} & \qty{30}{Hz} & \qty{0.1}{m}-\qty{2.4}{m} \\
    \qty{60}{Mhz} & \textit{Mode\_5\_60fps} & \qty{60}{Hz}& \qty{0.1}{m}-\qty{2.4}{m} \\

            \qty{80}{Mhz}-\qty{60}{Mhz} & \textit{Mode\_9\_15fps} & \qty{15}{Hz} & \qty{0.1}{m}-\qty{7.0}{m} \\
                \qty{80}{Mhz}-\qty{60}{Mhz} & \textit{Mode\_9\_20fps} & \qty{20}{Hz} & \qty{0.1}{m}-\qty{7.0}{m} \\
    \qty{80}{Mhz}-\qty{60}{Mhz} & \textit{Mode\_9\_30fps} & \qty{30}{Hz} & \qty{0.1}{m}-\qty{7.0}{m} \\

    \bottomrule
    \end{tabular}
    \end{center}
    \label{tab:flexx2_modes}
\end{table}

The PMD Flexx2\cite{pmdflexx2} is a 3D camera leveraging Time-of-Flight (ToF) technology to capture high-quality depth data.
It operates within a measurement range of \qty{0.1}{m} to \qty{7.0}{m}, with a specified depth resolution of $\leq$ 1\% of the measured distance.
Additionally, the PMD Flexx2 is designed to be energy-efficient, with a power consumption rate between \qty{570}{mW} to \qty{680}{mW}. It utilizes \qty{940}{nm} infrared light for illumination.
As outlined in \autoref{tab:camera_compare}, its compact and lightweight design combined with high FPS and long range makes it an ideal choice for mobile robotic applications, where environment perception is helpful for the control of robots.

The camera supports nine distinct operating modes. To ensure timely environment perception in robotic navigation, we selected modes that support frame rates of 30 FPS or higher: \textit{Mode\_5\_30fps}, \textit{Mode\_5\_60fps}, and \textit{Mode\_9\_30fps}, as detailed in \autoref{tab:flexx2_modes}. The operating modes starting with \textit{Mode 5} and \textit{Mode 9} are integrated with different filters to process the depth data by the manufacturer, while the last number indicates the sampling frequency.

\begin{figure}[t]
  \centering
  \subfloat[]{\includegraphics[width=0.5\linewidth]{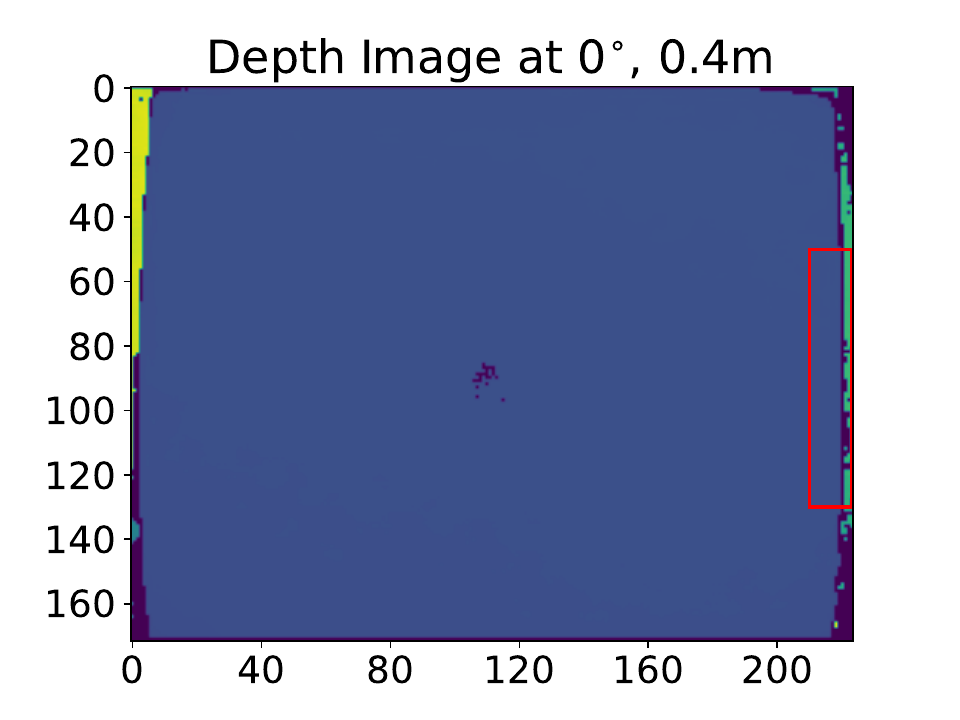}\label{fig:lateral_box_0}}
  \hfill
  \subfloat[]{\includegraphics[width=0.5\linewidth]{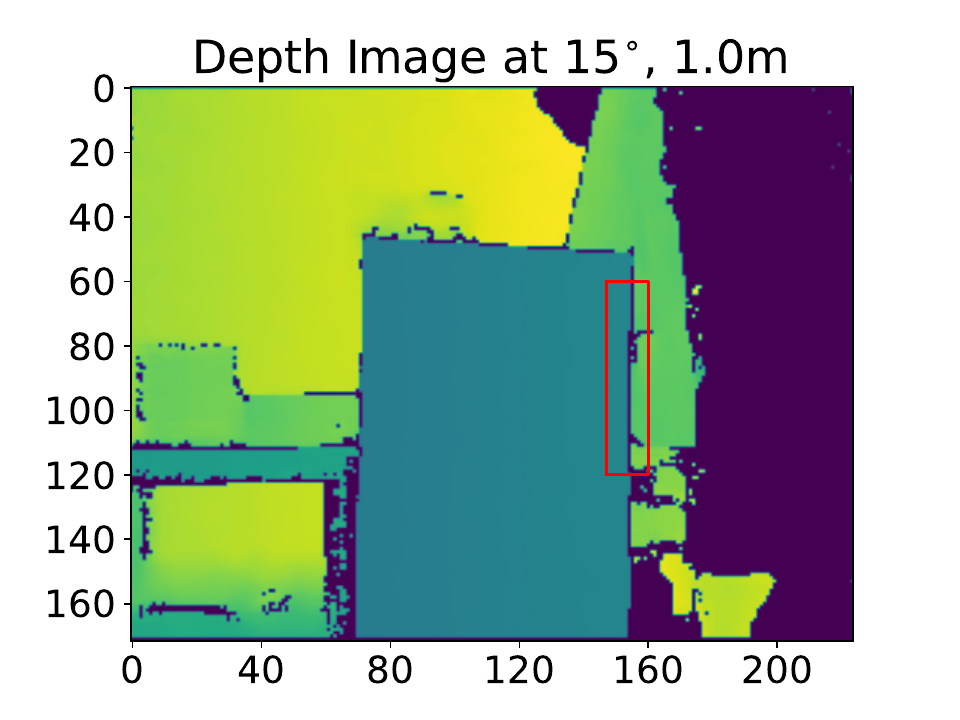}\label{fig:lateral_box_1}}
  \caption{Examples of depth images with region chosen for lateral noise analysis.}
  \label{fig:lateral_box}
\end{figure}

\subsection{Measurement Setup}\label{sec:measurement}

In our study, we conducted indoor experiments with the PMD Flexx2, adhering to methodologies outlined in \cite{khoshelham2012accuracy, nguyen2012modeling}. We selected a white flat surface on a painted wooden cabinet for these measurements, positioning the camera at various locations as illustrated in \autoref{fig:env_setup}, where $\mathbf{\Theta}$ indicates the incidence angle. We observed that the absorption and color of surface material can influence the results by causing more invalid pixels in the depth image, making the quantitative modeling of noise particularly challenging with black and metallic surfaces.
The cabinet remained stationary, while the camera was placed at specific distances and angles from the surface, with a precision of ±\qty{1}{cm}, achieved using measuring tapes. Following the coordinate conventions depicted in \autoref{fig:env_setup}, the camera's orientation was such that the Z-axis faced forward, and the positive Y-axis pointed downward. By adjusting the camera's location within the plane where both Z-axis and X-axis are located, we manipulated both the distance and the angle of incidence to the surface.
While the limited distance accuracy achieved with measuring tapes may introduce slight systematic errors, our analysis focuses on analyzing the camera's inherent non-systematic noise.

To gather depth data, we employed the Python API of the Royale SDK, provided by PMD.
For each set of conditions—operational mode, distance, and angle—we captured 300 frames of depth images without applying any further filtering.

\section{Depth Image Processing and Modeling Approach}\label{sec:modeling}

\begin{figure}[]
\setlength{\tabcolsep}{2pt} 
\begin{tabular}{cc}
 \includegraphics[width=0.48\linewidth]{{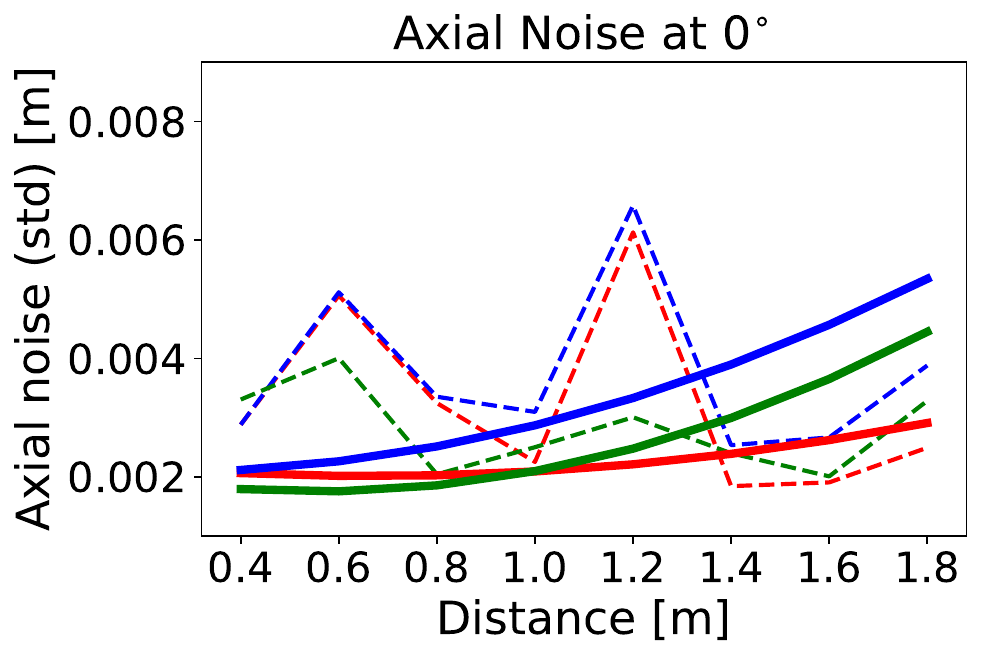}} &   
 \includegraphics[width=0.48\linewidth]{{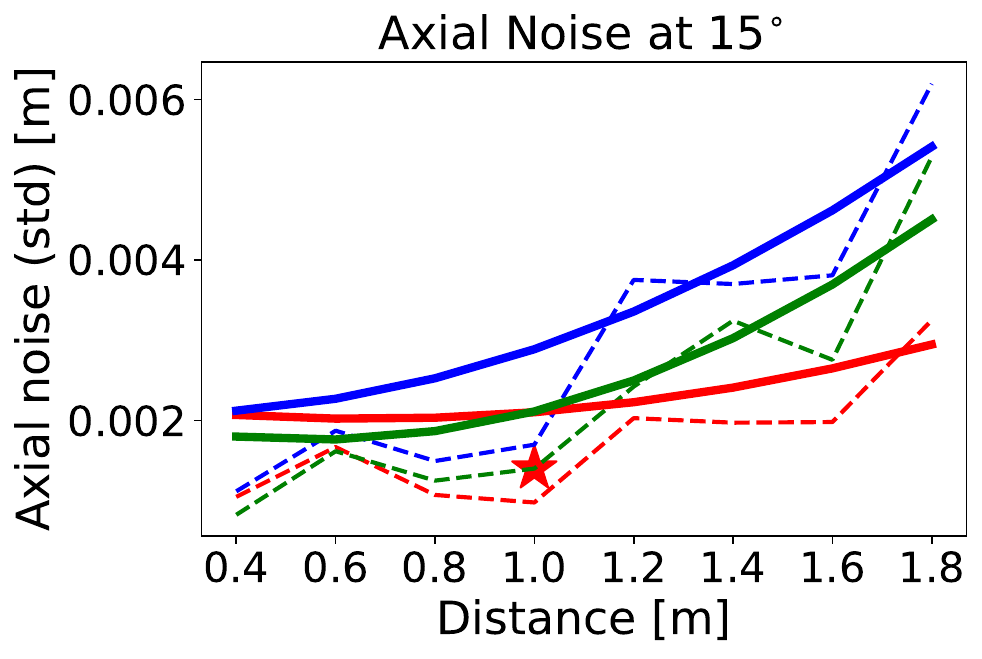}} \\
 \includegraphics[width=0.48\linewidth]{{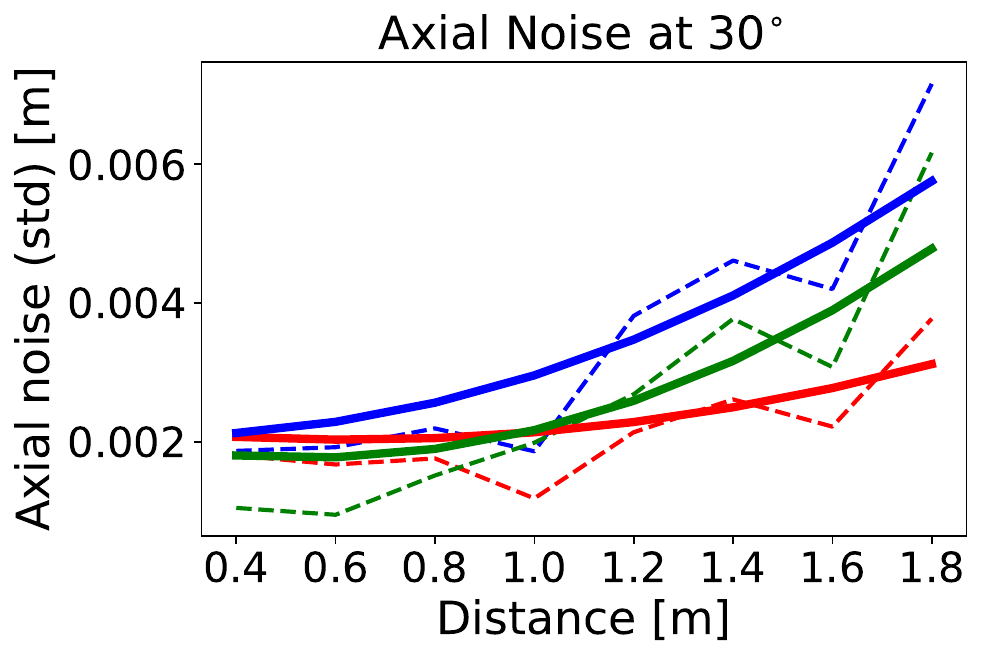}}  &
 \includegraphics[width=0.48\linewidth]{{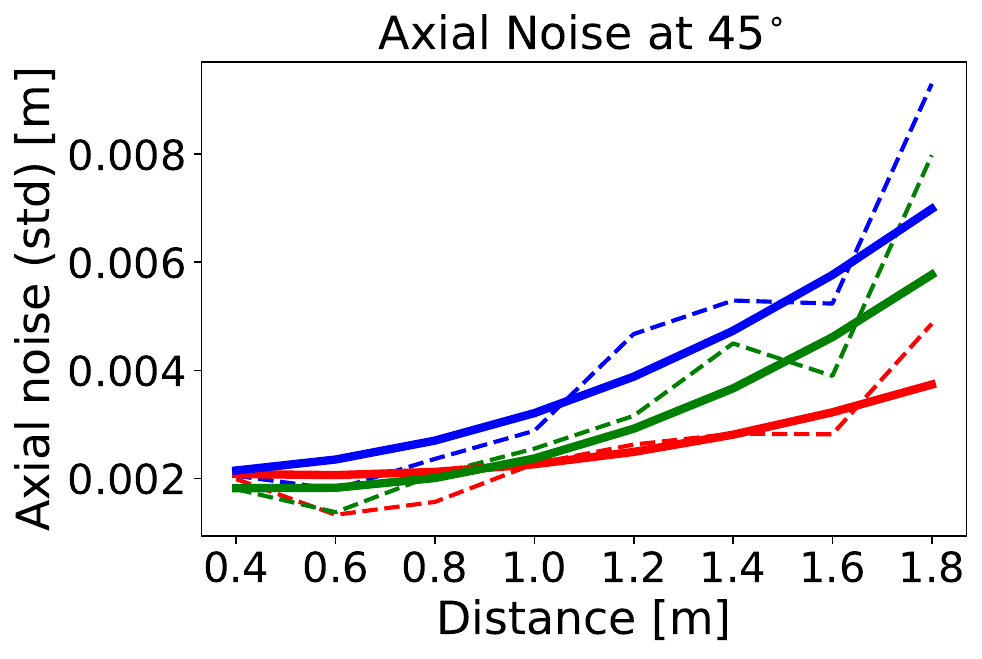}} \\  
 \includegraphics[width=0.48\linewidth]{{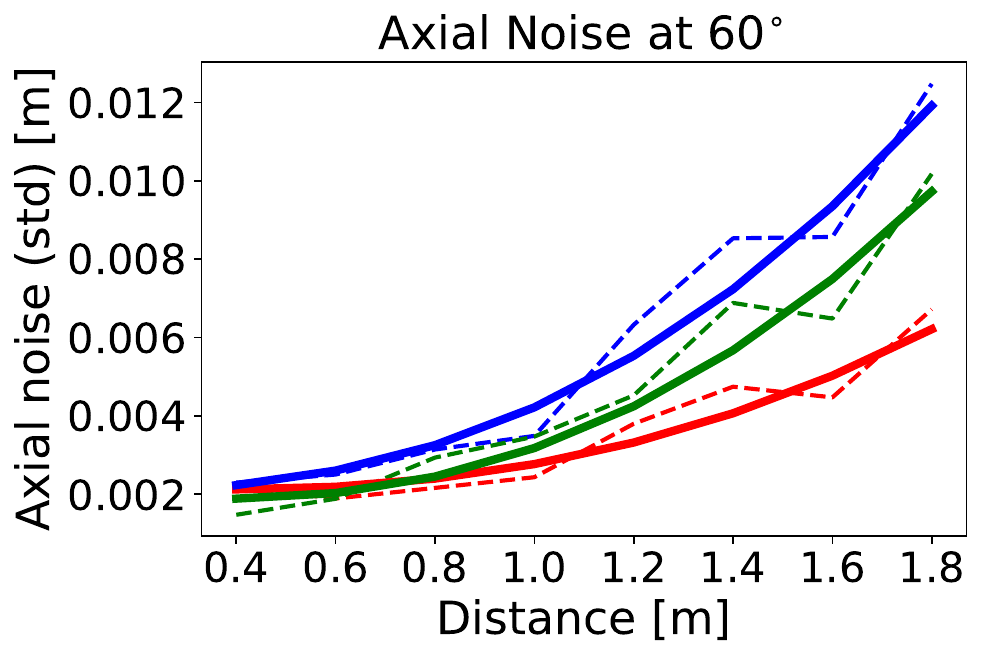}} & 
 \includegraphics[width=0.48\linewidth]{{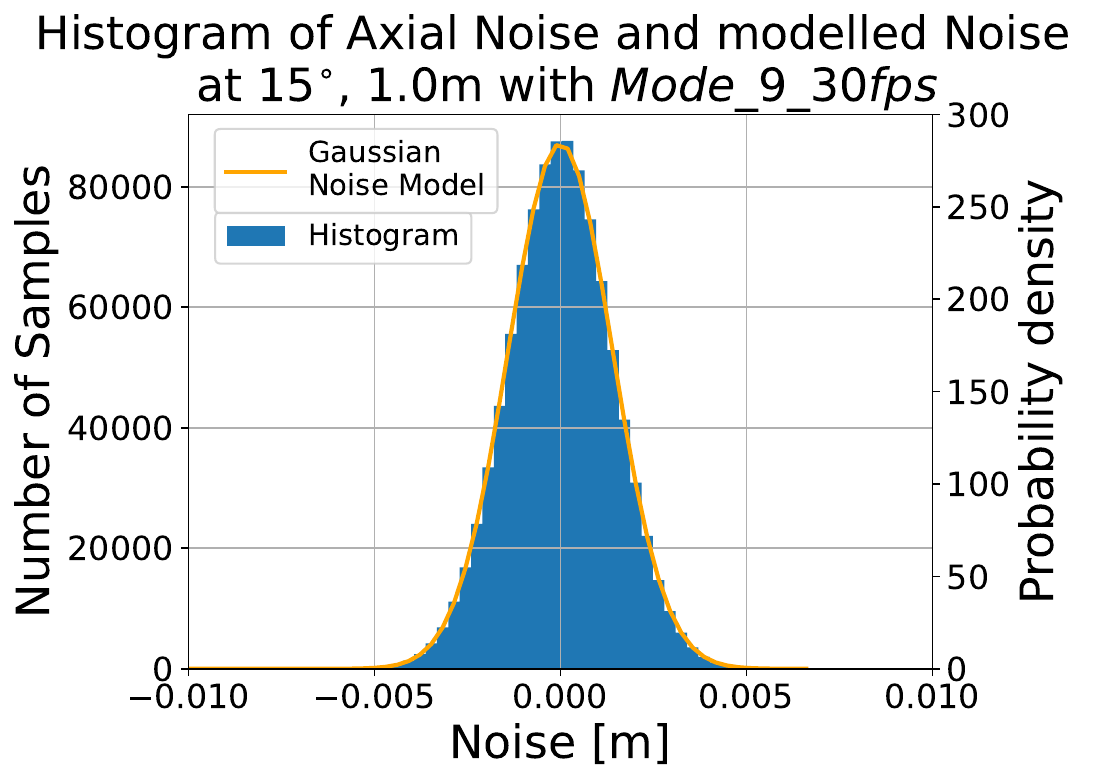}} \\
 \multicolumn{2}{c}{\includegraphics[height=0.7cm]{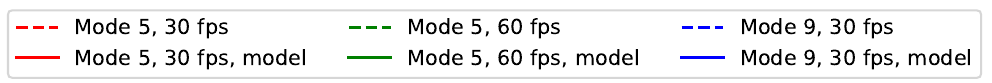} }\\
\end{tabular}
\caption{Standard deviation of measured data and model for axial noise at different incidence angles. The bottom right figure depicts the histogram of the axial noise for \textit{Mode\_9\_30fps} at incidence angle of 15$^{\circ}$ and distance \qty{1.0}{m} (marked in the upper right image) and the fitted Gaussian distribution.}
\label{fig:axial_noise_model}
\end{figure}

This paper applies a noise modeling approach inspired by similar work in other use-cases \cite{ahn2019analysis, fankhauser2015kinect, nguyen2012modeling} to analyze and model the noise characteristics of the PMD Flexx2. We measured and modeled the axial noise and lateral noise.
Axial noise refers to variations in depth measurements along the camera's line of sight, as depicted in \autoref{fig:axial_noise}, where $\mathbf{\Theta}$ is the incidence angle.  It causes the measured depth at a point to deviate from its true value, leading to inaccuracies in the depth data.
Lateral noise, in contrast, leads to discrepancies across the depth image, making the depth of one pixel inaccurately represent the depth of adjacent pixels. This effect can be seen in the jagged edges of objects in the depth map, where a smooth boundary is expected, as illustrated in \autoref{fig:lateral_noise}.

These noise types are characterized through a series of experiments that involve collecting depth images under varied environmental conditions, aiming to accurately fit corresponding noise models.
We model the distributions of both axial and lateral noise for each pixel as Gaussian distributions with independent and identically distributed values.

For axial noise analysis, we selected a depth image region within the flat plane that excludes the edges to minimize lateral noise interference, as illustrated in \autoref{fig:axial_box}.
This selection is crucial because including edge pixels in the analysis might inadvertently capture depth data from the background, distorting the axial noise evaluation.
Within this area, we fitted a plane to the data points for each set of conditions and then calculated the standard deviation of depth values over 300 frames, intentionally omitting any pixels that lacked depth information.
To model this noise, we adapted the axial noise models proposed in \cite{ahn2019analysis, nguyen2012modeling, fankhauser2015kinect} to align with the noise characteristics observed in the PMD Flexx2, as shown in \autoref{eq:axial_noise_fitting_formula} and discussed in detail in \autoref{sec:result}. This model was chosen since it was widely applied to different depth cameras, including Kinect V2, which operates on the same principle as the PMD Flexx2.

For lateral noise analysis, we chose a region adjacent to the plane's right vertical edge, as illustrated in \autoref{fig:lateral_box}. 
To quantify this noise, we computed the depth gradient along the X-axis for each row of the depth image, omitting any pixels with invalid data.
This computation aided in identifying the plane's left vertical edge by analyzing the gradient values
Subsequently, a vertical line was fitted to these identified edge pixels, and the standard deviation, measured in pixel units, was calculated to assess the lateral noise.

\section{Experimental Evaluation}\label{sec:result}

\begin{figure}[]
\captionsetup[subfloat]{farskip=-2pt,captionskip=1pt}
  \centering
  \subfloat[]{\includegraphics[width=0.5\linewidth]{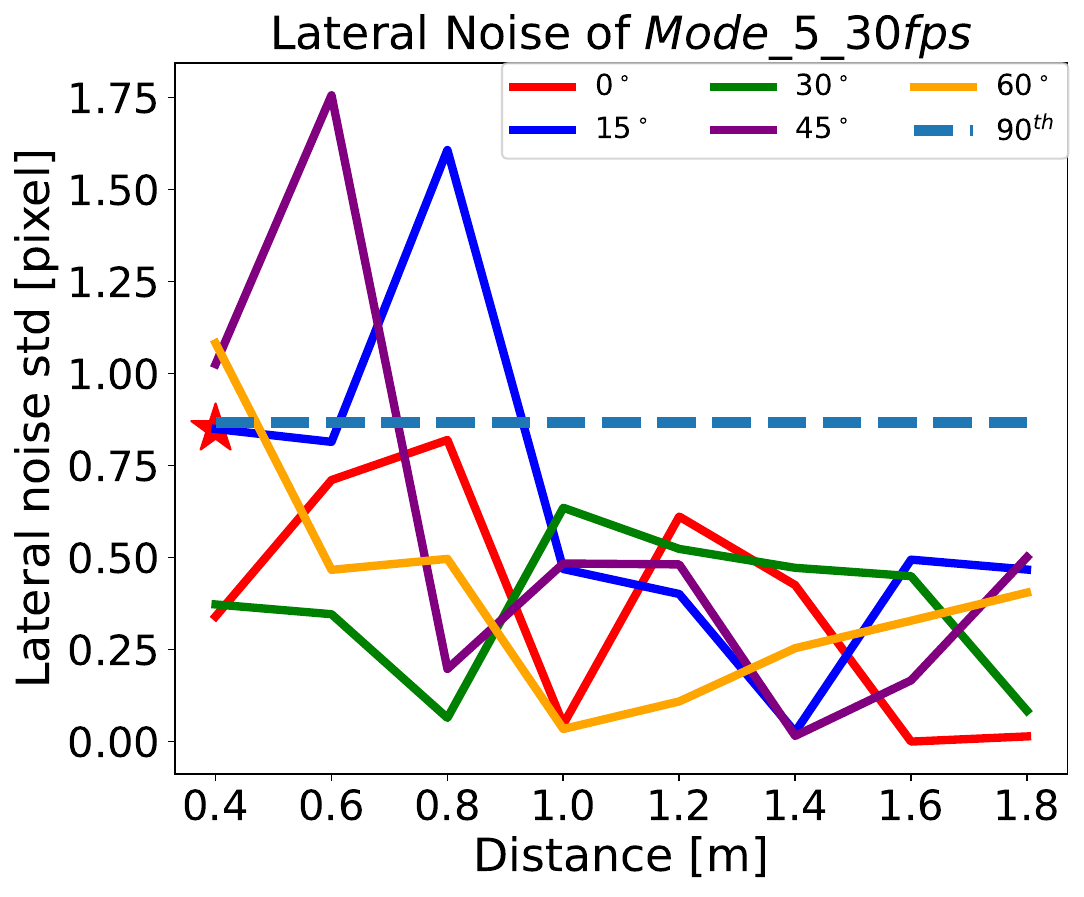}\label{fig:lateral_sub1}}
  \hfill
  \subfloat[]{\includegraphics[width=0.5\linewidth]{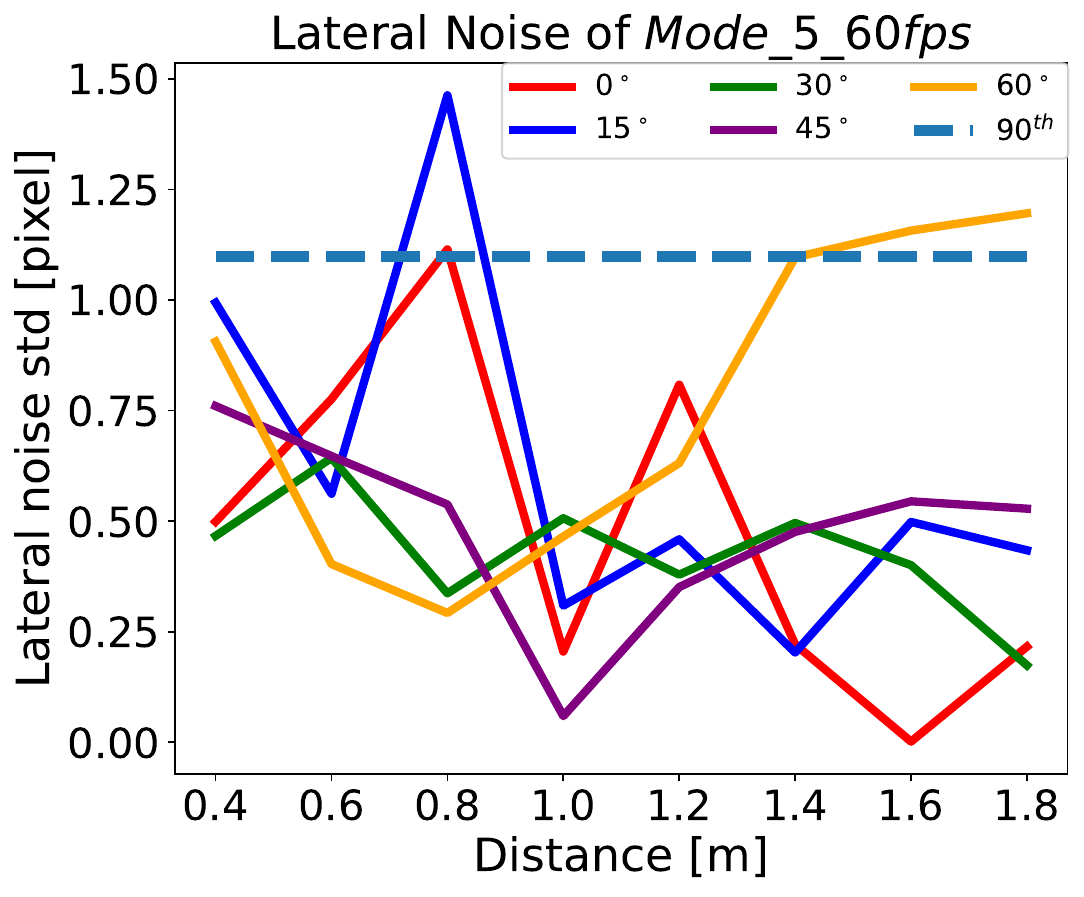}\label{fig:lateral_sub2}}
  \\
  \subfloat[]{\includegraphics[width=0.5\linewidth]{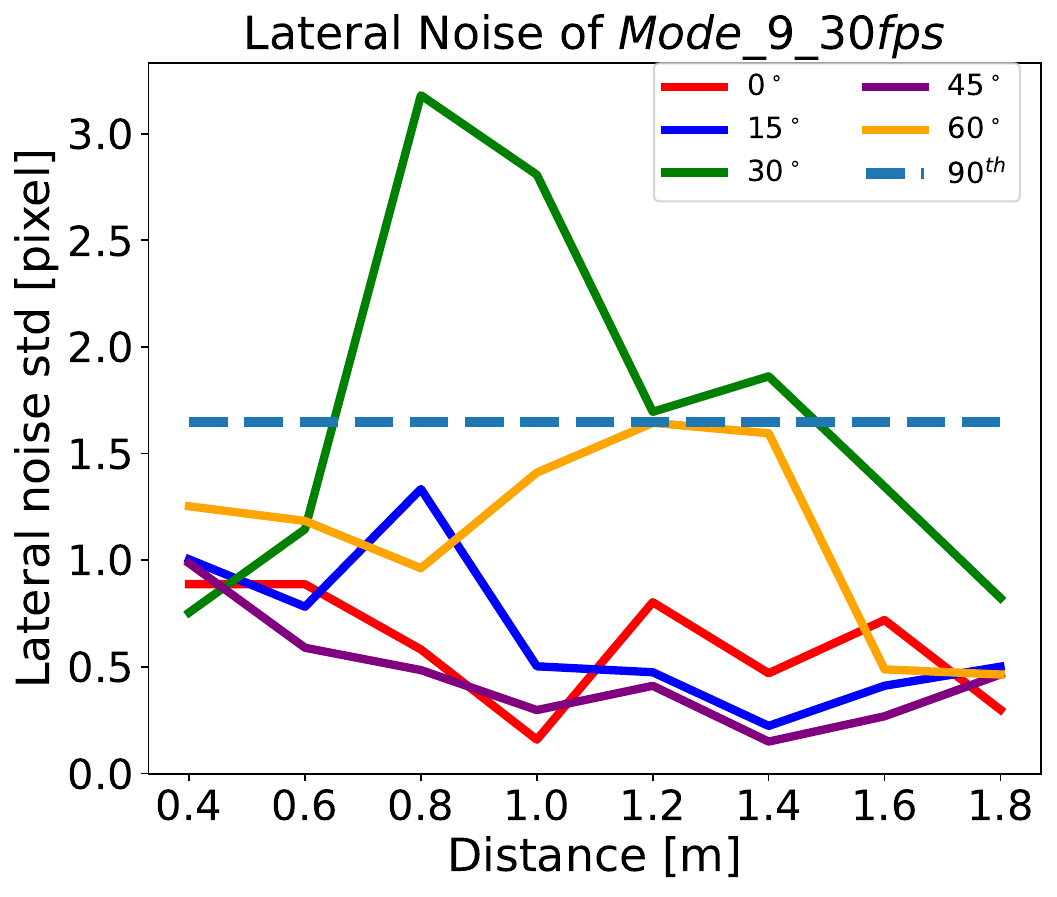}\label{fig:lateral_sub3}}
  \hfill
  \subfloat[]{\includegraphics[width=0.5\linewidth]{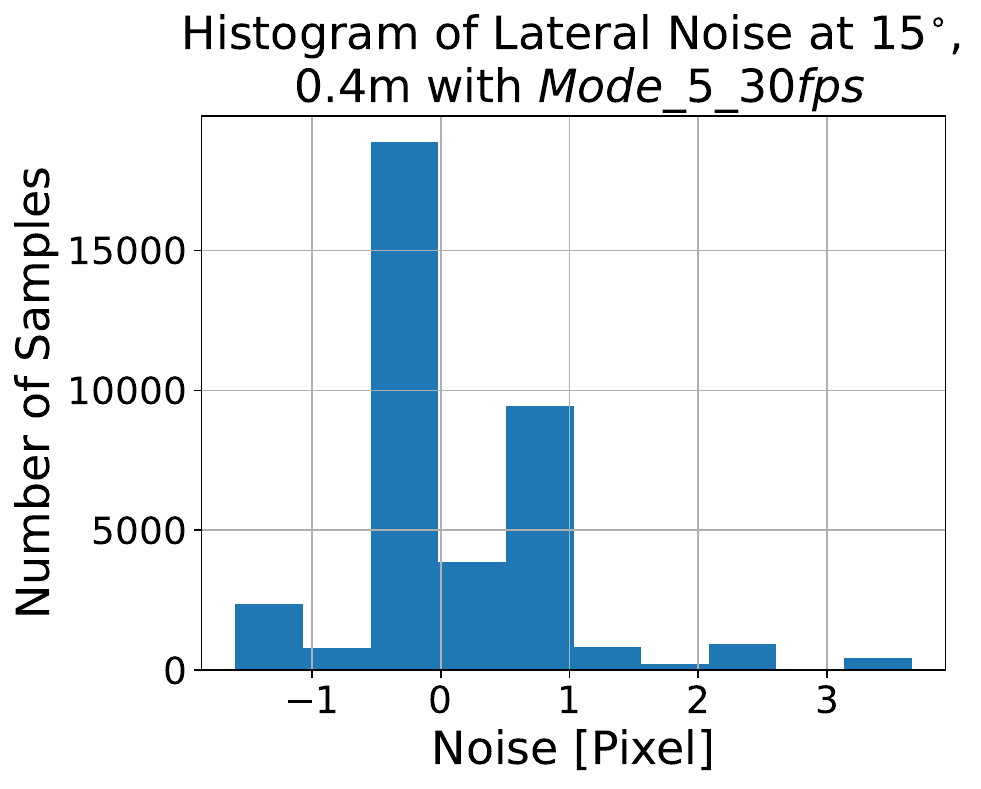}\label{fig:lateral_sub4}}
  \caption{ \ref{fig:lateral_sub1}-\ref{fig:lateral_sub3} Standard deviation of lateral noise with different modes,  \ref{fig:lateral_sub4} Distribution of lateral noise at \textit{mode\_5\_30fps}, incidence angle of 15$^{\circ}$ and distance \qty{0.4}{m} (marked in \autoref{fig:lateral_sub1}).}
  \label{fig:lateral_noise_model}
\end{figure}

The results of the axial and lateral noise measurements described in \autoref{sec:measurement} are presented in \autoref{fig:axial_noise_model} and \autoref{fig:lateral_noise_model} respectively.
We observed that \textit{Mode\_5\_30fps} typically exhibits the lowest noise levels among the three evaluated modes and is therefore suitable for applications where latency is less of a concern. Conversely, for scenarios where system latency is critical, \textit{Mode\_5\_60FPS} offers a viable solution. Additionally, \textit{Mode\_9\_30fps} can be considered for applications requiring long-range depth measurement.

\subsection{Axial Noise Model}

The characteristics of the measured axial noise are depicted in \autoref{fig:axial_noise_model}. Notably, the magnitude of axial noise increases with an increase in the distance from the camera when the incidence angle is equal to or exceeds 15 degrees. Furthermore, the axial noise follows a Gaussian distribution, as depicted in the bottom right figure of \autoref{fig:axial_noise_model}, in which the total number of samples is 1009800 and the number of bins is 100.
This observation underscores the similarity in axial noise behavior between the PMD Flexx2 and the Intel RealSense D435 cameras\cite{ahn2019analysis}. It is noteworthy that with incidence angles greater than 15$^{\circ}$, \textit{Mode\_5\_30fps} generally exhibits the lowest standard deviation in axial noise among the three tested modes. Conversely, when the camera is oriented directly towards the planar surface, \textit{Mode\_9\_30fps} demonstrates a reduced level of axial noise, suggesting its potential suitability for scenarios requiring direct camera alignment.
The anomaly at 0 degrees depicted in \autoref{fig:axial_noise_model} might be related to the principle of this camera (ToF) since in this condition specular reflections from the emitter can create sensor saturation problems, even with auto exposure enabled. For the noise modeling, we ignore this anomaly.

\begin{table}[b]
    \caption{Model coefficients for different modes.}
    \begin{center}
    \begin{tabular}{lccccc}
    \toprule
    \textbf{Coefficient} &  $\pmb{a}$   & $\pmb{b}$    &  $\pmb{c}$    &  $\pmb{d}$   &  $\pmb{n}$  \\
    \midrule
    \textit{Mode\_5\_30fps} & 0.002362 & -0.001041 & 0.000753 & 0.000185 & 2.7 \\
    \textit{Mode\_5\_60fps} & 0.002209 & -0.000793 & 0.001418 & 0.000370 & 2.7 \\
    \textit{Mode\_9\_30fps} & 0.002345 & -0.002101 & 0.001824 & 0.000298 & 2.7
    \end{tabular}
    \end{center}
    \label{tab:model_axial}
\end{table}

Our axial noise model, adapted from \cite{nguyen2012modeling}, is represented as a Gaussian distribution. The standard deviation of the noise $\sigma_z$, expressed in meters, is modeled by \autoref{eq:axial_noise_fitting_formula}, where $z$ is the depth in meters and $\theta_y$ is the angle of incidence.

\begin{equation}
    \begin{array}{r}
    \sigma_{z}(z, \theta_y) = a + b \cdot z + c \cdot z^2 + d \cdot z^{n}\frac{\theta_y^2}{(\pi/2 - \theta_y)^2}
    \end{array} 
    \label{eq:axial_noise_fitting_formula}
\end{equation}

By minimizing the mean squared error loss to fit the model to the measured data, we obtained the coefficients $a, b, c, d$.
In previous studies, values of $n$ were empirically chosen as 1.5 in \cite{fankhauser2015kinect} and -0.5 in \cite{khoshelham2012accuracy}.
In our study, we instead optimized $n$ for the Flexx2 by evaluating its fit to the data across a range from -1 to 3, in increments of 0.1, ultimately selecting 2.7 as the best fitting value.
The final models are visualized in \autoref{fig:axial_noise_model} and the coefficients are presented in \autoref{tab:model_axial}.

The quality of the resulting model is validated by computing the average pixel-wise KL divergence between the measured axial noise and the proposed model, a metric that has been used to validate RGB camera noise modeling \cite{abdelhamed2019noise}.
KL divergence was calculated for each combination of operational mode, distance, and incidence angle, using the selected pixels from all recorded images. 
\autoref{tab:kl_divergence_axial} details the average KL divergence values for various modes at different incidence angles, with an overall low average of \qty{0.015}{nats} across all modes.
Despite the previously discussed anomaly at 0 degrees, the results validate the overall good matching between our models and the measured data, indicating that the Gaussian model effectively captures the axial noise characteristics of the PMD Flexx2.

\begin{table}[t]
    \caption{Average KL divergence (in nats) between measured and modeled axial noise at different incidence angles.}
    \begin{center}
    \begin{tabular}{lccccc}
    \toprule
    \textbf{Incidence angle} & 0°    & 15°   & 30°   & 45°   & 60°   \\
    \midrule
    \textit{Mode\_5\_30fps} & 0.046 & 0.012 & 0.029 & 0.010 & 0.008 \\
    \textit{Mode\_5\_60fps} & 0.036 & 0.006 & 0.016 & 0.006 & 0.006 \\
    \textit{Mode\_9\_30fps} & 0.032 & 0.003 & 0.006 & 0.009 & 0.003
    \end{tabular}
    \end{center}
    \label{tab:kl_divergence_axial}
\end{table}

\subsection{Lateral Noise Model}
\autoref{fig:lateral_noise_model} illustrates the characteristics of the lateral noise obtained from the PMD Flexx2 camera. Unlike axial noise, lateral noise does not exhibit any recognizable trend, showing an absence of a clear relationship between the magnitude of noise and varying distances.
Although the distribution of lateral noise deviates from a Gaussian pattern as seen in \autoref{fig:lateral_sub4}, we still model it as such, following the approach described in \cite{ahn2019analysis}. We consider the noise to be independent of distance and incidence angle, conservatively setting the standard deviation $\sigma_x$ for each mode at the 90th percentile of the collected data, as shown in \autoref{fig:lateral_noise_model}.
The standard deviations for the different modes are detailed in \autoref{tab:model_lateral}.

\begin{table}[b]
   \caption{Standard deviation $\sigma_x$ of lateral noise of different modes }
   \begin{center}
   \begin{tabular}{lccc}
   \toprule
   \textbf{Mode}          & \textit{Mode\_5\_30fps} & \textit{Mode\_5\_60fps} & \textit{Mode\_9\_30fps} \\
   \midrule
       $\sigma_x$ [pixel]  & 0.864           & 1.098           & 1.649       
   \end{tabular}
   \end{center}
   \label{tab:model_lateral}
\end{table}

The model is validated using the same pixel-wise KL divergence metric as used for axial noise.
The resulting average KL divergences for the three modes \textit{Mode\_5\_30fps}, \textit{Mode\_5\_60fps} and  \textit{Mode\_9\_30fps} are \qty{0.896}{nats}, \qty{0.773}{nats}, and \qty{0.935}{nats} respectively, yielding an overall average of \qty{0.868}{nats}.
As expected, these higher values compared to axial noise suggest a less accurate Gaussian fit for lateral noise. However, the results are acceptably within a reasonable range, affirming our model as a good conservative estimate.

\section{Conclusion}\label{sec:conclu}

In conclusion, this paper provides a detailed characterization of the non-systematic noise of the PMD Flexx2 depth camera and introduces robust noise models for both axial and lateral distributions across various operating modes, specifically tailored for robotic applications.
Both noise types are assumed to follow Gaussian distributions. Axial noise is accurately modeled as a function of distance and incidence angle, achieving a high level of precision with a low average KL divergence of \qty{0.015}{nats} across all modes. Despite its inherent deviation from a Gaussian distribution, lateral noise is conservatively modeled, resulting in a satisfactory KL divergence of \qty{0.868}{nats}.

These findings affirm the effectiveness of our models, which are crucial for enhancing the simulation accuracy of depth cameras in virtual environments.
Such enhancements are expected to aid in the development and refinement of exteroceptive RL controllers for quadrupedal robotic locomotion. However, the precise impact of noise modeling on reducing the sim-to-real gap and improving the efficacy of RL-based controllers trained in simulations, ensuring their reliability in real-world applications, requires further validation.

While the proposed noise model for the PMD Flexx2 depth is effective, it is based on a single surface type. Future efforts will focus on diversifying surface materials and colors to refine and enhance the model's applicability. Additionally, implications of these enhancements on the simulation-to-reality gap will also be explored.

\section*{Acknowledgment}

This work was supported by the ETH Future Computing Laboratory (EFCL).

\balance    
\bibliographystyle{IEEEtran}
\bibliography{example}

\begin{thebibliography}{10}
\providecommand{\url}[1]{#1}
\csname url@samestyle\endcsname
\providecommand{\newblock}{\relax}
\providecommand{\bibinfo}[2]{#2}
\providecommand{\BIBentrySTDinterwordspacing}{\spaceskip=0pt\relax}
\providecommand{\BIBentryALTinterwordstretchfactor}{4}
\providecommand{\BIBentryALTinterwordspacing}{\spaceskip=\fontdimen2\font plus
\BIBentryALTinterwordstretchfactor\fontdimen3\font minus \fontdimen4\font\relax}
\providecommand{\BIBforeignlanguage}[2]{{%
\expandafter\ifx\csname l@#1\endcsname\relax
\typeout{** WARNING: IEEEtran.bst: No hyphenation pattern has been}%
\typeout{** loaded for the language `#1'. Using the pattern for}%
\typeout{** the default language instead.}%
\else
\language=\csname l@#1\endcsname
\fi
#2}}
\providecommand{\BIBdecl}{\relax}
\BIBdecl

\bibitem{mejia2019kinect}
J.~D. Mejia-Trujillo, Y.~J. Castano-Pino, A.~Navarro, J.~D. Arango-Paredes, D.~Rinc{\'o}n, J.~Valderrama, B.~Munoz, and J.~L. Orozco, ``Kinect™ and intel realsense™ d435 comparison: A preliminary study for motion analysis,'' in \emph{2019 IEEE International Conference on E-health Networking, Application \& Services (HealthCom)}.\hskip 1em plus 0.5em minus 0.4em\relax IEEE, 2019, pp. 1--4.

\bibitem{el2012study}
R.~A. El-laithy, J.~Huang, and M.~Yeh, ``Study on the use of microsoft kinect for robotics applications,'' in \emph{Proceedings of the 2012 IEEE/ION Position, Location and Navigation Symposium}.\hskip 1em plus 0.5em minus 0.4em\relax IEEE, 2012, pp. 1280--1288.

\bibitem{hutter2012starleth}
M.~Hutter, C.~Gehring, M.~Bloesch, M.~A. Hoepflinger, C.~D. Remy, and R.~Siegwart, ``Starleth: A compliant quadrupedal robot for fast, efficient, and versatile locomotion,'' in \emph{Adaptive mobile robotics}.\hskip 1em plus 0.5em minus 0.4em\relax World Scientific, 2012, pp. 483--490.

\bibitem{pinto2015evaluation}
A.~M. Pinto, P.~Costa, A.~P. Moreira, L.~F. Rocha, G.~Veiga, and E.~Moreira, ``Evaluation of depth sensors for robotic applications,'' in \emph{2015 IEEE international conference on autonomous robot systems and competitions}.\hskip 1em plus 0.5em minus 0.4em\relax IEEE, 2015, pp. 139--143.

\bibitem{zennaro2014evaluation}
S.~Zennaro, ``Evaluation of microsoft kinect 360 and microsoft kinect one for robotics and computer vision applications,'' 2014.

\bibitem{miki2022learning}
T.~Miki, J.~Lee, J.~Hwangbo, L.~Wellhausen, V.~Koltun, and M.~Hutter, ``Learning robust perceptive locomotion for quadrupedal robots in the wild,'' \emph{Science Robotics}, vol.~7, no.~62, p. eabk2822, 2022.

\bibitem{galna2014accuracy}
B.~Galna, G.~Barry, D.~Jackson, D.~Mhiripiri, P.~Olivier, and L.~Rochester, ``Accuracy of the microsoft kinect sensor for measuring movement in people with parkinson's disease,'' \emph{Gait \& posture}, vol.~39, no.~4, pp. 1062--1068, 2014.

\bibitem{funek2019evaluation}
G.~Halmetschlager-Funek, M.~Suchi, M.~Kampel, and M.~Vincze, ``An empirical evaluation of ten depth cameras: Bias, precision, lateral noise, different lighting conditions and materials, and multiple sensor setups in indoor environments,'' \emph{IEEE Robotics \& Automation Magazine}, vol.~26, no.~1, pp. 67--77, 2019.

\bibitem{ahn2019analysis}
M.~S. Ahn, H.~Chae, D.~Noh, H.~Nam, and D.~Hong, ``Analysis and noise modeling of the intel realsense d435 for mobile robots,'' in \emph{2019 16th International Conference on Ubiquitous Robots (UR)}.\hskip 1em plus 0.5em minus 0.4em\relax IEEE, 2019, pp. 707--711.

\bibitem{fankhauser2015kinect}
P.~Fankhauser, M.~Bloesch, D.~Rodriguez, R.~Kaestner, M.~Hutter, and R.~Siegwart, ``Kinect v2 for mobile robot navigation: Evaluation and modeling,'' in \emph{2015 international conference on advanced robotics (ICAR)}.\hskip 1em plus 0.5em minus 0.4em\relax IEEE, 2015, pp. 388--394.

\bibitem{khoshelham2012accuracy}
K.~Khoshelham and S.~O. Elberink, ``Accuracy and resolution of kinect depth data for indoor mapping applications,'' \emph{sensors}, vol.~12, no.~2, pp. 1437--1454, 2012.

\bibitem{wasenmuller2017comparison}
O.~Wasenm{\"u}ller and D.~Stricker, ``Comparison of kinect v1 and v2 depth images in terms of accuracy and precision,'' in \emph{Computer Vision--ACCV 2016 Workshops: ACCV 2016 International Workshops, Taipei, Taiwan, November 20-24, 2016, Revised Selected Papers, Part II 13}.\hskip 1em plus 0.5em minus 0.4em\relax Springer, 2017, pp. 34--45.

\bibitem{pasinetti2019performance}
S.~Pasinetti, M.~M. Hassan, J.~Eberhardt, M.~Lancini, F.~Docchio, and G.~Sansoni, ``Performance analysis of the pmd camboard picoflexx time-of-flight camera for markerless motion capture applications,'' \emph{IEEE Transactions on Instrumentation and Measurement}, vol.~68, no.~11, pp. 4456--4471, 2019.

\bibitem{pmdflexx2}
{PMD Technologies}, ``Pmd flexx2 3d time-of-flight camera,'' Available: \url{https://3d.pmdtec.com/en/3d-cameras/flexx2/}, 2024, [Online; accessed 12-April-2024].

\bibitem{haider2022cameranoise}
\BIBentryALTinterwordspacing
A.~Haider and H.~Hel-Or, ``What can we learn from depth camera sensor noise?'' \emph{Sensors}, vol.~22, no.~14, 2022. [Online]. Available: \url{https://www.mdpi.com/1424-8220/22/14/5448}
\BIBentrySTDinterwordspacing

\bibitem{yan2020depth}
C.~Yan, Z.~Li, Y.~Zhang, Y.~Liu, X.~Ji, and Y.~Zhang, ``Depth image denoising using nuclear norm and learning graph model,'' \emph{ACM Transactions on Multimedia Computing, Communications, and Applications (TOMM)}, vol.~16, no.~4, pp. 1--17, 2020.

\bibitem{mallick2014characterizations}
T.~Mallick, P.~P. Das, and A.~K. Majumdar, ``Characterizations of noise in kinect depth images: A review,'' \emph{IEEE Sensors journal}, vol.~14, no.~6, pp. 1731--1740, 2014.

\bibitem{nguyen2012modeling}
C.~V. Nguyen, S.~Izadi, and D.~Lovell, ``Modeling kinect sensor noise for improved 3d reconstruction and tracking,'' in \emph{2012 second international conference on 3D imaging, modeling, processing, visualization \& transmission}.\hskip 1em plus 0.5em minus 0.4em\relax IEEE, 2012, pp. 524--530.

\bibitem{margolis2023walk}
G.~B. Margolis and P.~Agrawal, ``Walk these ways: Tuning robot control for generalization with multiplicity of behavior,'' in \emph{Conference on Robot Learning}.\hskip 1em plus 0.5em minus 0.4em\relax PMLR, 2023, pp. 22--31.

\bibitem{lee2020learning}
J.~Lee, J.~Hwangbo, L.~Wellhausen, V.~Koltun, and M.~Hutter, ``Learning quadrupedal locomotion over challenging terrain,'' \emph{Science robotics}, vol.~5, no.~47, p. eabc5986, 2020.

\bibitem{gervet2023navigating}
T.~Gervet, S.~Chintala, D.~Batra, J.~Malik, and D.~S. Chaplot, ``Navigating to objects in the real world,'' \emph{Science Robotics}, vol.~8, no.~79, p. eadf6991, 2023.

\bibitem{breuer2014low}
T.~Breuer, C.~Bodensteiner, and M.~Arens, ``Low-cost commodity depth sensor comparison and accuracy analysis,'' in \emph{Electro-Optical Remote Sensing, Photonic Technologies, and Applications VIII; and Military Applications in Hyperspectral Imaging and High Spatial Resolution Sensing II}, vol. 9250.\hskip 1em plus 0.5em minus 0.4em\relax Spie, 2014, pp. 77--86.

\bibitem{lachat2015first}
E.~Lachat, H.~Macher, M.-A. Mittet, T.~Landes, and P.~Grussenmeyer, ``First experiences with kinect v2 sensor for close range 3d modelling,'' in \emph{ICIAP 2015 Workshops: BioFor, CTMR, RHEUMA, ISCA, MADiMa, SBMI, and QoEM (2015-09-07 to 2015-09-08: Genoa, Italy)}, vol.~40, 2015.

\bibitem{giancola2018metrological}
S.~Giancola, M.~Valenti, R.~Sala, S.~Giancola, M.~Valenti, and R.~Sala, ``metrological qualification of the intel d400™ active stereoscopy cameras,'' \emph{A Survey on 3D Cameras: Metrological Comparison of Time-of-Flight, Structured-Light and Active Stereoscopy Technologies}, pp. 71--85, 2018.

\bibitem{bahnsen20213d}
C.~H. Bahnsen, A.~S. Johansen, M.~P. Philipsen, J.~W. Henriksen, K.~Nasrollahi, and T.~B. Moeslund, ``3d sensors for sewer inspection: A quantitative review and analysis,'' \emph{Sensors}, vol.~21, no.~7, p. 2553, 2021.

\bibitem{condotta2020evaluation}
I.~C. Condotta, T.~M. Brown-Brandl, S.~K. Pitla, J.~P. Stinn, and K.~O. Silva-Miranda, ``Evaluation of low-cost depth cameras for agricultural applications,'' \emph{Computers and Electronics in Agriculture}, vol. 173, p. 105394, 2020.

\bibitem{abdelhamed2019noise}
A.~Abdelhamed, M.~A. Brubaker, and M.~S. Brown, ``Noise flow: Noise modeling with conditional normalizing flows,'' in \emph{Proceedings of the IEEE/CVF International Conference on Computer Vision}, 2019, pp. 3165--3173.

\end{thebibliography}

\end{document}